\documentclass[intlimits,twoside,a4paper]{article}

\usepackage{amsmath,amssymb}
\usepackage{graphicx}
\usepackage{wrapfig}
\usepackage[T2A]{fontenc}
\usepackage[cp1251]{inputenc}

\usepackage{cmpj2}

\articletype{Rapid Communication}

\issue{2015}{18}{2}{24601}
\doinumber{10.5488/CMP.18.24601}
\title[Gas-liquid critical point of the ultrasoft restricted primitive model]%
{Gas-liquid critical point of the ultrasoft restricted primitive model from analytic theory}
\author[O. Patsahan]{O. Patsahan}

\address{Institute for Condensed Matter Physics of the National
Academy of Sciences of Ukraine, \\1 Svientsitskii St., 79011 Lviv,
Ukraine}

\date{Received April 22, 2015}

\begin{document}

\maketitle

\begin{abstract}
Gas-liquid criticality in the ultrasoft restricted primitive model (URPM) of polyelectrolytes is studied using the collective
variables-based theory. For the model, an  effective Hamiltonian is derived and explicit expressions for all the coefficients are
found in a one-loop approximation. Based on this Hamiltonian, the phase and critical behaviour is analysed.
Our results provide evidence that the nature of the gas-liquid criticality  in the URPM is the same as in the restricted primitive
model that includes a hard core.

\keywords polyelectrolytes,  ultrasoft restricted primitive model, gas-liquid critical point, effective Hamiltonian
\pacs 64.75.-g, 64.60.De, 64.75.Cd, 64.60.fd
\end{abstract}

Recently, a  new model of  polyelectrolyte solutions, the so-called ultrasoft restricted primitive model (URPM), has been introduced
in \cite{Coslovich_Hansen_Kahl:11-1,Coslovich_Hansen_Kahl:11-2}. The URPM is a globally  electroneutral  equimolar mixture of
positive and negative spatially extended charge distributions. Contrary to the familiar restricted primitive model (RPM)
\cite{hansen_mcdonald}, no hard cores are
involved in the URPM.  Nevertheless, the URPM admits a thermodynamic limit since it is H-stable in the sence of Fisher and
Ruelle \cite{Ruelle,Fisher_Ruelle}. The phase diagram of the model was studied  in
\cite{Coslovich_Hansen_Kahl:11-1,Coslovich_Hansen_Kahl:11-2,Caillol__Levesque:14}
using  Monte Carlo (MC) and molecular dynamic  simulations.  The analysis of simulation data for relatively small system sizes
revealed a nontrivial topology of the phase diagram: a region of low temperature  gas-liquid phase coexistence  terminating at a
critical point  and  a sharp conductor-insulator transition  above this point \cite{Coslovich_Hansen_Kahl:11-1,Coslovich_Hansen_Kahl:11-2}.  The simulation data also hinted at  a tricritical
nature of the gas-liquid critical point. Such a phase behaviour   qualitatively differs from that found in the RPM.
More recently \cite{Caillol__Levesque:14}, the gas-liquid phase transition and the critical behavior of the URPM have
been studied using  finite size scaling MC simulations in the grand canonical ensemble.
While these simulation results confirm the existence of the gas-liquid critical point,  they fail to provide a conclusive evidence for   the nature of criticality of the model
as well as  reliable values for the  critical exponents. In addition, the estimates of the critical point parameters ($T_{\mathrm{c}}^{*}=0.0134$ and $\rho_{\mathrm{c}}^{*}=0.26$)
differ from those given in \cite{Coslovich_Hansen_Kahl:11-2}, i.e., $T_{\mathrm{c}}^{*}\sim 0.018$, $\rho_{\mathrm{c}}^{*}\sim 0.16$
(expressed in the same reduced units as in \cite{Caillol__Levesque:14}) which are obtained for smaller system sizes and
without finite scaling size analysis. In general, the simulation results   show a strong dependence of the coexistence envelope
(its location and shape) on the system size  compared to the case of the RPM.
The gas-liquid phase coexistence in the URPM has been also predicted by the mean-field (MF) like theories
\cite{Nikoubashman__Hansen_Kahl:12,Warren__Masters:13}, although  with the  critical point being
considerably distant from the  simulations. As expected,  the MF  theories predict  a classical critical behavior near
the critical point.

Motivated by the above mentioned simulation studies,  we address the issue of  the gas-liquid criticality in the
URPM using the theory  that
exploits the method of
collective variables (CVs) \cite{Yuk-Hol,Pat-Mryg-CM}. The
theory enables us to derive, on the microscopic grounds, the
 effective Hamiltonian and  find  all the
relevant coefficients, including the square-gradient term,  within
the framework  of the same approximation \cite{Patsahan:13}.
Using this theory, we  obtained the  effective Hamiltonian of the RPM in the vicinity of the gas-liquid critical point  that takes into account the
contribution from higher order correlations between the positive and
negative ions \cite{Patsahan:13,patsahan:04:1}.  The resulting form of this  Hamiltonian  confirms the fact that its critical behaviour   belongs  to the universal class of a three-dimensional Ising model.
Herein below, we derive the effective Hamiltonian of the URPM and, on this basis, analyse the phase and critical behaviour.

We consider an equimolar mixture of $N_{+}$ polycations of charge $+Q$ and $N_{-}$ ($=N_{+}$) polyanions of charge $-Q$ in a
volume $V$ with the total number density $\rho=N/V$ ($N=N_{+}+N_{-}$).  The domain of  the volume $V$ occupied by polyions is denoted by $\Omega$. Polyions bear an extended charge distribution $\pm Q\rho(r)$
where the normalized distribution $\rho(r)$ is supposed to be the same for both species and is given by  a Gaussian law
\begin{displaymath}
\rho(r)=\frac{\exp[-r^{2}/(2\bar\sigma^{2})]}{(2\pi\bar\sigma^{2})^{3/2}}\,,
\end{displaymath}
$\bar{\sigma}=\sigma/2$ is the radius  of the polyion. The interaction potential  between   a
polyion of species $\alpha$ and one of the species $\beta$ is as follows \cite{Coslovich_Hansen_Kahl:11-2}:
\begin{equation}
 u_{\alpha\beta}(r)=\frac{Q_{\alpha}Q_{\beta}}{r}{{\rm{erf}}}(r/2\bar{\sigma}).
 \label{pot_urpm}
\end{equation}
This  pair potential is finite at a full overlap [${\rm{erf}}(r/2\bar{\sigma})\sim r$ as $r\rightarrow 0$] ensuring the existence of
the thermodynamic limit \cite{Caillol__Levesque:14,Ruelle,Fisher_Ruelle}. The Fourier transform of $ u_{\alpha\beta}(r)$ has the form:
\begin{equation}
 \tilde{u}_{\alpha\beta}(k)=\frac{4\pi Q_{\alpha}Q_{\beta}}{k^{2}}\exp(-k^{2}\bar{\sigma}^{2}).
 \label{furier_urpm}
\end{equation}

The model (\ref{pot_urpm}) is at equilibrium in the grand canonical ensemble,  and  by $\beta=1/k_{\mathrm{B}}T$ we denote the inverse
temperature ($k_{\mathrm{B}}$ is the Boltzmann constant) and $\mu$ is the chemical potential. We use the method of CVs,  with
the ideal gas system being a  reference system. In this case, one can use  the exact functional representation  of the grand
partition function derived for a  multicomponent  system  in  \cite{Patsahan_Mryglod:12} (see equation~(18)
in \cite{Patsahan_Mryglod:12}). For the model (\ref{pot_urpm}), it can be presented as follows:
\begin{equation}
\Xi=\int \mathcal{D}\rho_{N} \int \mathcal{D}\rho_{Q}\; \exp
\left(- {\cal H}[\rho_{N},\rho_{Q}]\right), \label{VSS_sad}
\end{equation}
where the Hamiltonian ${\cal H}[\rho_{N},\rho_{Q}]$ has the form:
\begin{eqnarray}
{\cal H}[\rho_{N},\rho_{Q}]&=&-\int {\rm d} {\mathbf
r}\,\rho_{N}(\mathbf{r})+
\frac{1}{2}\int {\rm d} {\mathbf
r}\,\left[\rho_{N}(\mathbf{r})+\rho_{Q}(\mathbf{r})\right]\ln\left[\frac{\rho_{N}(\mathbf{r})+\rho_{Q}(\mathbf{r})}{2}\right]
\nonumber\\
&&{}
+\frac{1}{2}\int {\rm d} {\mathbf
r}\,\left[\rho_{N}(\mathbf{r})-\rho_{Q}(\mathbf{r})\right]
\ln\left[\frac{\rho_{N}(\mathbf{r})-\rho_{Q}(\mathbf{r})}{2}\right]
+\frac{\beta}{2}\int {\rm d} {\mathbf r} {\rm d} {\mathbf r}'\;
u_{QQ}(|{\mathbf r}- {\mathbf r}'|)
\nonumber\\
&&{}
\times
\rho_{Q}(\mathbf{r})\rho_{Q}(\mathbf{r}')
-\int {\rm d} {\mathbf
r}\,\rho_{N}(\mathbf{r})\bar{\nu}_{N}\,. \label{action_FT}
\end{eqnarray}
\looseness=-1In the above equations, $\rho_{N}$ and $\rho_{Q}$ denote the two CVs:  $\rho_{N}({\mathbf r})=\rho_{+}({\mathbf r})+\rho_{-}({\mathbf r})$
describing the field of the total number  density and  $\rho_{Q}({\mathbf r})=\rho_{+}({\mathbf r})-\rho_{-}({\mathbf r})$  describing the field of the charge density ($\rho_{+(-)}({\mathbf r})$, in turn, describes the field of the number density of the cations (anions)).
$u_{QQ}$ is a linear combination of the interaction potentials $ u_{\alpha\beta}(r)$:
$u_{QQ}(r)=\frac{1}{4}\left[u_{++}(r)-2u_{+-}(r)+u_{--}(r)\right]$.
 $\bar{\nu}_{N}=(\bar{\nu}_{+}+\bar{\nu}_{-})/2$, where $\bar\nu_{\alpha}({\mathbf
r})=\nu_{\alpha}+\nu_{\alpha}^{S}$ is the dimensionless chemical potential of the $\alpha$th
species, $\nu_{\alpha}=\beta\mu_{\alpha}-3\ln\Lambda_{\alpha}$,
$\Lambda_{\alpha}$ is
the de Broglie thermal wavelength and $\nu_{\alpha}^{S}$ is
the self-energy of the $\alpha$th species $\nu_{\alpha}^{S}=\beta u_{\alpha\alpha}({\mathbf r},{\mathbf r})/2=\beta u_{0}/2$ with $u_{0}=Q^{2}/(\sqrt{\pi}\bar{\sigma})$. For the model under consideration,
$u_{QQ}(r)=u_{++}(r)=u_{--}(r)=|u_{+-}(r)|$
and $\bar{\nu}_{N}=\bar{\nu}_{+}=\bar{\nu}_{-}=\bar{\nu}$.

Functional integrals (\ref{VSS_sad})--(\ref{action_FT})  can be given a precise meaning in the case where
the domain $\Omega$ is a cube of side $L$ ($V=L^{3}$) with periodic
boundary conditions. This means that we restrict ourselves to the fields $\rho_{A}(\mathbf{r})$ ($A=N,Q$)
which can be written as Fourier series
$\rho_{A}(\mathbf{r})=V^{-1}\sum_{\mathbf{k}\in \Lambda}\rho_{\mathbf{k},A} \exp(\rm i\mathbf{k}\mathbf{r})$,
where $\Lambda=(2\pi/L)\mathbb{Z}^{3}$ is the reciprocal cubic lattice \cite{Patsahan_Mryglod_Caillol:07,Goldenfeld}.

Expanding the entropic part of Hamiltonian (\ref{action_FT}) in powers of $\rho_{N}$ and $\rho_{Q}$ (more exactly, in powers of deviations
of $\rho_{N}$ and $\rho_{Q}$ from their MF values), we arrive at the  expression similar
to that  obtained in \cite{Patsahan:13} (see equation (7) in \cite{Patsahan:13}). The main difference  is
that in the case of the URPM, the contributions from the hard sphere reference system to  series expansion coefficients $a_{n}^{(i_{n})}$   reduce to
the ideal gas terms.
Next, following the procedure developed in \cite{Patsahan:13}, one can derive the effective Hamiltonian of the URPM
near the gas-liquid critical point.
Here, we  directly use the expressions obtained in  \cite{Patsahan:13} replacing the contributions from the hard sphere system
with the corresponding terms for the ideal gas model. As a result,  the effective Hamiltonian in the Fourier
space representation reads as
\begin{eqnarray}
{\cal H}^{\mathrm{eff}}&=&a_{1,0}\rho_{0,N}+\frac{1}{2!\langle
N\rangle}\sum_{{\mathbf{k}}}\left(a_{2,0}+k^{2}a_{2,2}\right)\rho_{{\bf
k},N}\rho_{-{\bf k},N}
\nonumber \\
&&{}+\sum_{n\geqslant  3}\frac{1}{n!\langle
N\rangle^{n-1}}\sum_{{\mathbf{k}}_{1},\ldots,{\mathbf{k}}_{n}}a_{n,0}\rho_{{\bf
k_{1}},N}\rho_{{\bf k_{2}},N}\ldots\rho_{{\bf
k_{n}},N}\delta_{{\bf{k}}_{1}+\ldots +{\bf{k}}_{n}}\,,
\label{H_eff}
\end{eqnarray}
where $\delta_{{\bf{k}}_{1}+\ldots +{\bf{k}}_{n}}$ is the Kronecker symbol. The coefficients in a one-loop approximation have the form:
\begin{eqnarray}
a_{1,0} &=& -\Delta\nu_{N}-\frac{1}{2\langle
N\rangle}\sum_{\mathbf{q}}\widetilde g_{Q}(q), \label{a10}\\
a_{n,0} &=&(-1)^{n}(n-2)!-\frac{(n-1)!}{2\langle
N\rangle}\sum_{\mathbf{q}}\left[\widetilde g_{Q}(q)\right]^{n}, \qquad  n\geqslant  2,
\label{an0} \\
a_{2,2} &=&-\frac{1}{4\langle
N\rangle}\sum_{\mathbf{q}}\widetilde g_{Q}^{(2)}(q)\left[1+\widetilde
g_{Q}(q)\right],
\label{a22}
\end{eqnarray}
where  $\Delta\nu_{N}=\bar\nu_{N}-\nu_{N}^{{\mathrm{id}}}$, the superscript ``id'' refers to the ideal gas system. $\widetilde g_{Q}(q)$ is the screened potential
\begin{equation}
\widetilde g_{Q}(q)=-\frac{\beta\rho \widetilde u_{QQ}(q)}{1+\beta\rho
\widetilde u_{QQ}(q)}
\label{g_Q}
\end{equation}
and
the superscript $(2)$ in
equation~(\ref{a22}) denotes the second-order derivative of $\widetilde g_{Q}(q)$ with respect to the wave vector.
The terms involving  screened potentials  arise  from  integration  in (\ref{VSS_sad}) over the charge subsystem
(CVs  $\rho_{Q}$).

Taking into account (\ref{furier_urpm}) and (\ref{g_Q}), one can  obtain
 explicit expressions for coefficients (\ref{a10})--(\ref{a22}):
\begin{eqnarray}
  a_{1,0} &=&-\nu+\nu^{\mathrm{id}}-\frac{1}{2T^{*}}+\frac{1}{\sqrt{\pi}}\int_{0}^{\infty}\frac{x^{2}{\rm  d}x}{T^{*}x^{2}e^{x^{2}}+24\eta\sqrt{\pi}}\,,
 \label{a10_ex}   \\
a_{n,0} &=&(-1)^{n}(n-2)!\left[1-(n-1)\int_{0}^{\infty}\frac{(24\eta)^{n-1}\sqrt{\pi}^{n-2}x^{2}{\rm  d}x}{(T^{*}x^{2}e^{x^{2}}+24\eta\sqrt{\pi})^{n}} \right],
\qquad n\geqslant  2,
\label{an0_ex}\\
a_{2,2} &=& -48\eta\sqrt{\pi}{T^{*}}^{2}\int_{0}^{\infty}\frac{x^{4}e^{2x^{2}}}{(T^{*}x^{2}e^{x^{2}}+24\eta\sqrt{\pi})^{4}}\big[24\eta\sqrt{\pi}
  (3+7x^{2}+2x^{4}) \nonumber \\
 &&-T^{*}x^{2}e^{x^{2}}(1+x^{2}+2x^{4})\big]{\rm  d}x.
 \label{a22_ex}
\end{eqnarray}
In equations~(\ref{a10_ex})--(\ref{a22_ex}),   the following notations are introduced:  $T^{*}=k_{\mathrm{B}}T/u_{0}=k_{\mathrm{B}}T\sqrt{\pi}\bar{\sigma}/Q^{2}$,
$\eta=\pi\rho\bar{\sigma}^{3}/6$,
and  $x=k\bar{\sigma}$. Coefficient $a_{1,0}$ is the excess part of the chemical potential
$\nu$, and the equation $a_{1,0}=0$ yields  the chemical potential in
the random phase approximation (RPA).

It should be emphasized that although the original
Hamiltonian of the URPM [equation~(\ref{action_FT})] does not include direct
pair attractive  interactions of total number density fluctuations, the
effective short-range attraction  does appear in the effective
Hamiltonian. This attractive interaction is the result of taking into account
the charge-charge correlations through integration over the charge
subsystem. Therefore, one can state that the mechanism of the gas-liquid separation in the URPM is the same as in the RPM.
 The form of the effective Hamiltonian (\ref{H_eff})--(\ref{a22}) suggests  the Ising-like critical behaviour of the URPM.
Herein below, we use explicit expressions~(\ref{a10_ex})--(\ref{a22_ex}) for the
analysis  of the gas-liquid phase separation.

First, we consider the critical point. At the critical point, the  system of equations
 \begin{eqnarray}
  a_{2,0}(\rho_{\mathrm{c}},T_{\mathrm{c}})=0, \qquad a_{3,0}(\rho_{\mathrm{c}},T_{\mathrm{c}}) =0
\label{cr-point}
 \end{eqnarray}
holds.   In addition to equations (\ref{cr-point}), we solve the equation $a_{4,0}=0$.
The loci of the above-mentioned equations in the ($T^{*},\rho^{*}$) plane are shown in figure~\ref{fig1}. The solid line corresponding to the solutions of the first
equation in (\ref{cr-point})  is called a spinodal line. The two lines (solid and dashed)  intersect at a maximum of the gas-liquid
spinodal yielding the coordinates of the critical point. As is seen, coefficient $a_{4,0}$ does not vanish at the critical point suggesting that this point
is a regular critical point.

 \begin{figure}[h]
 \centering
 \includegraphics[height=5.5cm]{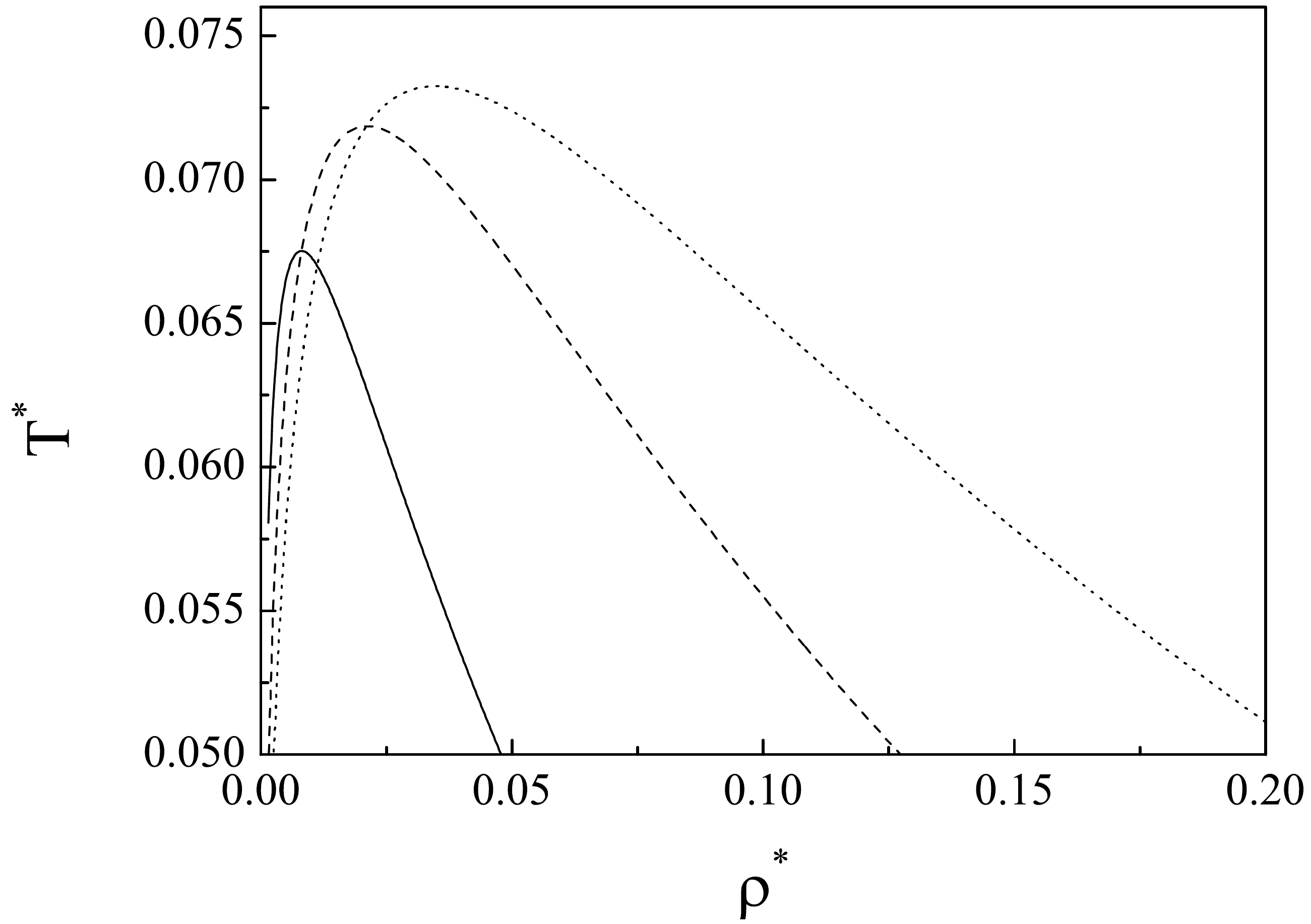}
 \caption{Loci of equations $a_{2,0}=0$
 (solid line),  $a_{3,0}=0$ (dashed line), and $a_{4,0}=0$ (dotted line). Temperature and density are given in
 dimensionless unites: $T^{*}=k_{\mathrm{B}}T\sqrt{\pi}\sigma/(2Q^{2})$  and
 $\rho^{*}=\rho\sigma^{3}$.} \label{fig1}
 \end{figure}
Critical temperature and critical density expressed in reduced units are presented in
table~\ref{tab1}. In order  to make some contact with the results obtained for the RPM, hereafter we use the same reduced units
for the density as in \cite{Caillol__Levesque:14}, i.e., $\rho^{*}=\rho\sigma^{3}$ where $\sigma$ is the  diameter of the polyion.  Our choice of   the reduced  temperature, $T^{*}=k_{\mathrm{B}}T/u_{0}$
with $u_{0}$  being the maximum strength of the attractive interaction, coincides  with that of
\cite{Coslovich_Hansen_Kahl:11-1,Coslovich_Hansen_Kahl:11-2,Caillol__Levesque:14,Nikoubashman__Hansen_Kahl:12}. In particular, one gets
 $u_{0}=2Q^{2}/(\sqrt{\pi}\sigma)$ for the URPM and $u_{0}=Q^{2}/\sigma$  in the case of
the RPM where $\sigma$ is the  diameter of the polyion in the former case and the  diameter of the hard sphere/ion in the latter case. With the critical parameters defined in such a way,  the critical temperature of the URPM appears to be smaller than the critical
temperature  of the RPM obtained in the same approximation, whereas  the critical densities of the both models
are nearly the same. As expected,  our estimates of the critical parameters of  the URPM coincide with the results  obtained
in  \cite{Warren__Masters:13} in the RPA. It should be noted that the RPA, like other mean-field theories
\cite{Nikoubashman__Hansen_Kahl:12},  predicts a far too
high critical temperature
and a far too low critical density compared  with the available simulation data \cite{Coslovich_Hansen_Kahl:11-1,Coslovich_Hansen_Kahl:11-2,Caillol__Levesque:14}.
Some possible reasons for such a situation have been discussed in \cite{Nikoubashman__Hansen_Kahl:12}.

\begin{table}[htbp]
\caption{Reduced  gas-liquid critical parameters, coefficients of the
effective  Hamiltonian and the  reduced Ginzburg temperature $t_{\mathrm{G}}$ for the ultrasoft restricted primitive model (URPM) and for the restricted primitive model (RPM)   in
the one-loop approximation. Temperature and density are given in
 dimensionless unites: $T_{\mathrm{c}}^{*}=k_{\mathrm{B}}T_{\mathrm{c}}/u_{0}$  and
 $\rho_{\mathrm{c}}^{*}=\rho_{\mathrm{c}}\sigma^{3}$ (see the text for details). The data for the RPM are taken from \cite{Patsahan:13}.}
\vspace{2ex}
\begin{center}
\begin{tabular}{|c|c|c|c|c|c|c|}
 \hline\hspace{5mm}
 Model\hspace{6mm} &\hspace{6mm}
$T_{\mathrm{c}}^{*}$\hspace{6mm} &\hspace{2mm}
$\rho_{\mathrm{c}}^{*}$\hspace{6mm}&\hspace{2mm}
$a_{2,t}$\hspace{6mm}&\hspace{2mm}
$a_{2,2}/\sigma^{2}$\hspace{6mm}&\hspace{2mm} $a_{4,0}$\hspace{8mm}&\hspace{5mm}
$t_{\mathrm{G}}$\hspace{5mm}
\\
\hline\hline
URPM&  $0.06750$    &$0.0081$  &   $1.0005$  &  $0.2693$  &   $0.1384$     & $0.0031$\\
RPM & $0.08446$ &$0.0088$&  $1.0758$ & $0.2570$ & $0.1752$  & $0.0053$  \\
\hline
\end{tabular}
\label{tab1}
\end{center}
\end{table}
To calculate the coexistence curve, first,   equating the right-hand side of (\ref{a10_ex}) to zero, we obtain an expression for
the chemical potential. Then, the Maxwell double-tangent construction is employed.  Figure~\ref{fig2} shows the coexistence curve of the
URPM (a solid line) in the ($T^{*}$,$\rho^{*}$) plane.  The coexistence curve of the RPM (a dashed line) is presented for comparison.
As is seen, the shape of the both curves is very similar. The inset shows the both coexistence curves in the corresponding-state variables:
$t^{*}=(T^{*}-T_{\mathrm{c}}^{*})/T_{\mathrm{c}}^{*}$ and $\Delta \rho^{*}=(\rho^{*}-\rho_{\mathrm{c}}^{*})/\rho_{\mathrm{c}}^{*}$.  Comparing  the
two coexistence envelopes in this case, one can see that  the URPM has a slightly wider coexistence region than the RPM.

 Finally,   using equations (\ref{an0_ex}) and (\ref{a22_ex}), we calculate  the reduced Ginzburg temperature $t_{\mathrm{G}}$ where
$t_{\mathrm{G}}=(T_{\mathrm{G}}-T_{\mathrm{c}})/T_{\mathrm{c}}$ ($T_{\mathrm{G}}$ is the Ginzburg temperature and $T_{\mathrm{c}}$ is the MF critical temperature).
According to the Ginzburg criterion \cite{Goldenfeld},  $t_{\mathrm{G}}$ marks a lower
 bound of the temperature region where   a MF description is self-consistent.
 For $|t|\ll t_{\mathrm{G}}$, Ising critical behavior should be exhibited.
One expects a crossover from  the MF behaviour to the Ising-like
critical behaviour when the reduced temperature $|t|$ becomes of the order of $t_{\mathrm{G}}$ \cite{Chaikin_Lubensky}.
The reduced Ginzburg temperature can be expressed in
terms of  coefficients of the Hamiltonian (\ref{H_eff})  as follows \cite{fisher3}:
\begin{eqnarray}
 t_{\mathrm{G}}=\displaystyle\frac{1}{32\pi^{2}}\frac{a_{4,0}^{2}}{a_{2,t}
 a_{2,2}^{3}}\,,
\label{t_G}
\end{eqnarray}
where $a_{2,t}=\left.\partial a_{2,0}/\partial t\right|_{t=0}$. Taking into account (\ref{an0_ex}), one can
obtain for $a_{2,t}$
 \begin{eqnarray}
 a_{2,t}=48\eta T_{\mathrm{c}}^{*}\int_{0}^{\infty}\frac{x^{4}e^{x^{2}}{\rm
 d}x}{(T_{\mathrm{c}}^{*}x^{2}e^{x^{2}}+24\eta\sqrt{\pi})^{3}}\,,
 \label{a2t_urpm}
 \end{eqnarray}
 where the same  dimensionless units as in (\ref{an0_ex}) are  used.

 \begin{figure}[htbp]
 \centering
 \includegraphics[height=5.5cm]{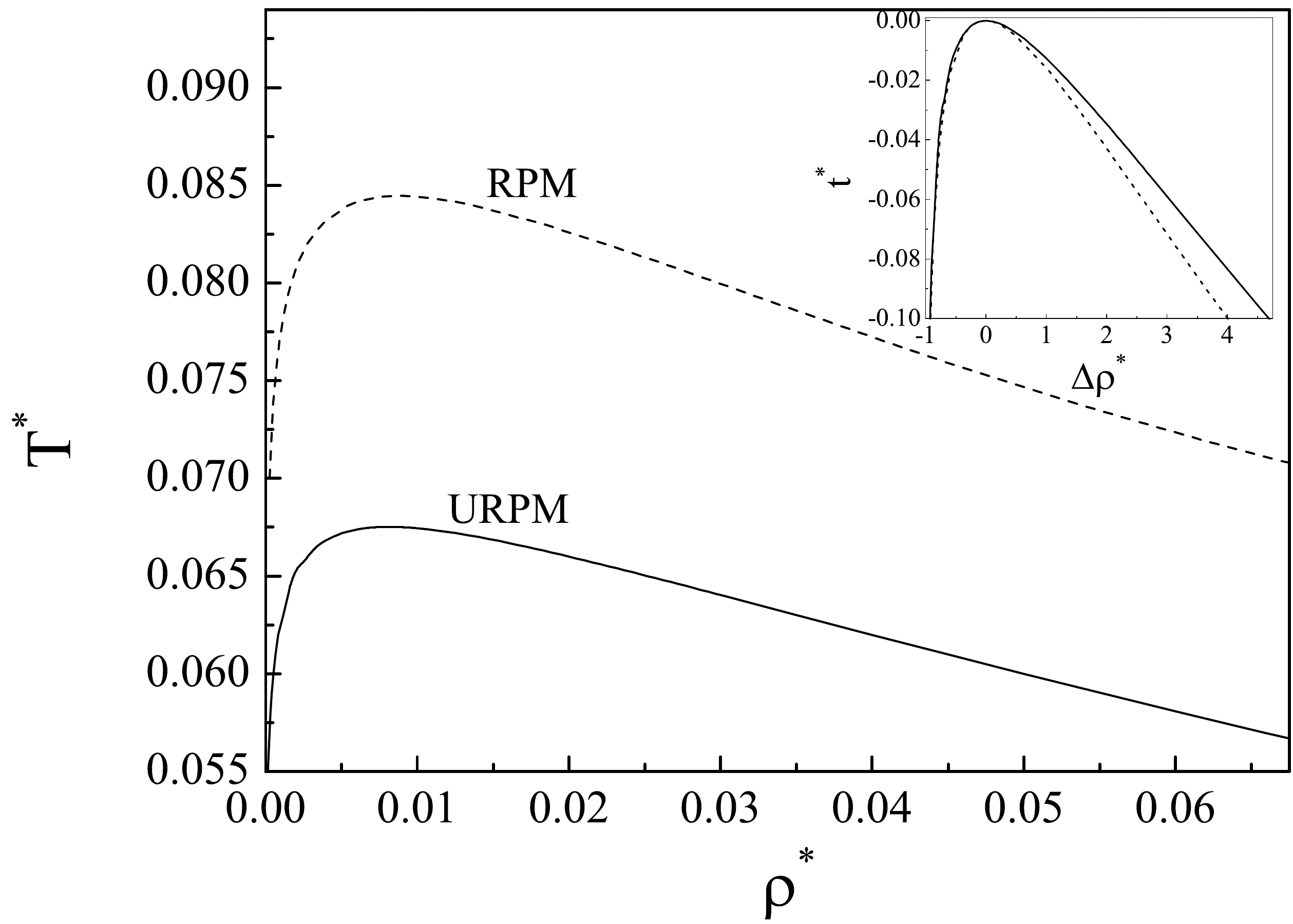}
 \caption{Gas-liquid coexistence  curve  of the URPM   in the ($T^{*}$,$\rho^{*}$)
 representation. The coexistence curve of the RPM (a dashed line) is presented for comparison (see the text for details). The inset shows
 the both coexistence curves in the
 corresponding-state variables.
 } \label{fig2}
 \end{figure}
Substituting the parameters of the  URPM critical point
 in equations (\ref{an0_ex}) (for $n=4$), (\ref{a22_ex}) and (\ref{t_G})--(\ref{a2t_urpm}), one gets the values for the Hamiltonian coefficients
 and the reduced Ginzburg temperature in the one-loop approximation. The results are presented  in table~\ref{tab1}.
The corresponding values  for the RPM obtained within the framework of the same approximation in \cite{Patsahan:13} are
presented in
the table for comparison. As is seen, there is only a little difference between the results obtained for the both models.
In particular, the models have much the same  region of the effective density-density attractions: $\sqrt{a_{2,2}}/\sigma\simeq 0.5$.
It is essential   that the reduced Ginzburg temperature obtained
for the URPM is close to that found for the RPM, i.e., $t_{\mathrm{G}}^{\mathrm{URPM}}\simeq 0.6 t_{\mathrm{G}}^{\mathrm{RPM}}$.
It is worth noting that  the reduced Ginzburg temperature of the RPM is
about $20$ times smaller than for the purely nonionic  model (a one-component hard-sphere square-well model) \cite{Patsahan:13}.

In summary,  we have derived  a microscopic-based effective Hamiltonian
of the URPM in the vicinity of the gas-liquid critical point and have found analytic expressions for all
the  coefficients  in the one-loop approximation. As for the RPM, the form of the effective
Hamiltonian  indicates that the  critical behavior of the URPM  belongs to the universal class of a
three-dimensional Ising model.
Based on this Hamiltonian, we  have calculated the reduced Ginzburg temperature $t_{\mathrm{G}}$   which, according to the Ginzburg criterion,
marks the onset  of the critical region.
We have found that  $t_{\mathrm{G}}$ for the URPM is  of the same order  as for the RPM.
In addition, our results have  shown that  the  shapes of  coexistence envelope  of  the URPM and the RPM
resemble each other when presented in the similar dimensionless units.
Our results suggest that   the gas-liquid  criticality in   both models, i.e., the URPM and the RPM, should be of the same nature.

\ukrainianpart
 \title{Критична точка газ-рідина ультрам'якої обмеженої примітивної моделі:  аналітична теорія}
 \author{О. Пацаган}
 \address{Інститут фізики конденсованих систем НАН України, вул. Свєнціцького, 1, 79011 м. Львів, Україна}
%
 \makeukrtitle
 \begin{abstract}
 Вивчається  критичність газ-рідина ультрам'якої обмеженої примітивної моделі (URPM) поліелектролітів, використовуючи теорію,
 що базується на методі колективних змінних. Для цієї моделі виведено ефективний гамільтоніан і отримано явні вирази для всіх
 коефіцієнтів в однопетлевому наближенні. На основі цього гамільтоніану аналізується фазова і критична поведінка. Отримані
 результати свідчать про те, що природа критичності газ-рідина в URPM є такою ж, як і в обмеженій примітивній моделі, яка включає
 твердий кор.
 \keywords поліелектроліти, ультрам'яка обмежена примітивна модель,  критична точка газ-рідина, ефективний гамільтоніан
 \end{abstract}

\end{document}